\documentclass{aastex63}

\usepackage[utf8]{inputenc}
\DeclareUnicodeCharacter{0301}{*************************************}

\newcommand{\dv}[2]{\frac{\mathrm{d} #1}{\mathrm{d} #2}}

\usepackage{hyperref}

\shorttitle{Spectral Line Depth Variability}
\shortauthors{Wise et al.}

\begin{document}

\title{Spectral Line Depth Variability in Radial Velocity Spectra}

\correspondingauthor{Alexander Wise}
\email{aw@psu.edu}

\author[0000-0002-5013-5769]{Alexander Wise}
\affiliation{Department of Astronomy and Astrophysics, The Pennsylvania State University, University Park, PA 16802, USA}
\affiliation{Center for Exoplanet and Habitable Worlds, The Pennsylvania State University, University Park, PA 16802, USA}

\author[0000-0002-8864-1667]{Peter Plavchan}
\affiliation{Department of Physics and Astronomy, George Mason University, Fairfax, VA 22030, USA}

\author[0000-0002-9332-2011]{Xavier Dumusque}
\affiliation{Astronomy Department, University of Geneva, 51 Chemin des Maillettes, 1290 Versoix, Switzerland}

\author[0000-0001-8934-7315]{Heather Cegla}
\affiliation{Physics Department, University of Warwick, Coventry CV4 7AL, United Kingdom}
\affiliation{Astronomy Department, University of Geneva, 51 Chemin des Maillettes, 1290 Versoix, Switzerland}

\author[0000-0001-7294-5386]{Duncan Wright}
\affiliation{University of Southern Queensland, Centre for Astrophysics, Australia}
\begin{abstract}

Stellar active regions, including spots and faculae, can create radial velocity (RV) signals that interfere with the detection and mass measurements of low mass exoplanets. In doing so, these active regions affect each spectral line differently, but the origin of these differences is not fully understood. Here we explore how spectral line variability correlated with S-index (Ca H \& K emission) is related to the atomic properties of each spectral line. Next we develop a simple analytic stellar atmosphere model that can account for the largest sources of line variability with S-index. Then we apply this model to HARPS spectra of $\alpha$~Cen~B to explain Fe I line depth changes in terms of a disk-averaged temperature difference between active and quiet regions on the visible hemisphere of the star. This work helps establish a physical basis for understanding how stellar activity manifests differently in each spectral line, and may help future work mitigating the impact of stellar activity on exoplanet RV surveys.

\end{abstract}


\section{Introduction}

The radial velocity (RV) method has been a prolific method of confirming and discovering exoplanets in the past two decades (see e.g. \cite{wright17} for a review). State-of-the-art
Extreme Precision Radial Velocity (EPRV)
instruments have instrumental precision of 20-30 cm/s \citep{pepe13, jurgenson16, blackman20, suarez20}, but stellar activity generates RV signals with amplitude 0.5-5 m/s in typical exoplanet survey stars \citep[e.g.][]{dumusque11}. Stellar activity is defined here as any spatial or temporal variability in brightness or velocity on the visible stellar disk that can manifest as a detectable RV signal. Some types of stellar activity occur on timescales that make averaging them down with many observations unfeasible (see e.g. \cite{plavchan20} for a summary). Granulation and super-granulation vary on timescales of minutes to hours \citep{derosa04,brandt08,miklos20} and magnetic activity such as spots, faculae and plage vary on timescales of
hours to years
\citep{baliunas95,saar97,haywood20}. To improve RV precision beyond the current limits set by stellar activity, the imprint of stellar activity on RVs and their underlying spectra must be understood.

Whole-spectrum cross-correlation function (CCF) line shape indicators have been used to diagnose and characterize stellar activity \citep[e.g.][]{deBeurs20,collier21}. However, several studies have looked at individual spectral line variability to find out how stellar activity affects each line differently. \cite{thompson17} have found for $\alpha$~Cen~B, and now the Sun \citep{thompson20}, that the depth of some spectral lines were significantly correlated with
emission in the cores of
the Ca II H \& K lines. Similarly, \cite{wise18} (hereafter W18) used the same data from $\alpha$~Cen~B to measure correlation coefficients between spectral line shapes and the Ca
II H \& K
lines. \cite{dumusque18} found that the RVs of some spectral lines were much more affected by stellar activity than others, and by curating the spectral lines on which the stellar RV is calculated, it is possible to mitigate stellar activity by a factor of two. Finally, \cite{cretignier20} pushed further the analysis performed in \cite{dumusque18} and found that the lines the most affected by stellar activity were the shallow ones, formed deep inside the stellar atmosphere, where convection is strong and therefore is strongly affected by the magnetic field responsible for stellar activity. Line-by-line analysis seems to be a promising tool to help us understand and model stellar activity \citep[e.g.][]{cretignier2021}.

Here we push further the results of W18 by developing an astrophysical toy model of their observed spectral line variability. W18 analyze High Accuracy Radial velocity Planet Searcher (HARPS) spectra of $\alpha$~Cen~B during a period of high magnetic activity in 2010 identified in \cite{dumusque12}. They quantify activity-induced line variability as the Kendall's Tau correlation between the Mt. Wilson S-index \citep{wilson78} and each line's depth, width, or center of mass.
The S-index measures Ca H \& K emission and is sensitive to magnetic activity.
W18 found that generally line depth variations had the most significant correlations with S-index. Our primary aim is to develop an astrophysical model to explain these most significant spectral line variations with activity.
We also investigate relationships between correlation coefficients measured in W18 and several atomic properties of spectral lines.

The paper is organized as follows.
In Section~\ref{sec:correlations}, we investigate relationships between the atomic properties of spectral lines and correlations linking spectral line measurements to S-index.
In Section~\ref{sec:model}, we develop an analytic model to explain the interaction between line core flux vs. S-index correlation coefficients and excitation energy.
Finally, in Section~\ref{sec:conclusion}, we summarize our findings and discuss future work.

\section{Correlations with Atomic Properties} \label{sec:correlations}

Here we search for correlations between activity-related line variations and atomic properties of spectral lines.
These properties can tell us how lines change in response to variations in the solar surface, such as temperature or magnetic field strength changes.
We use correlation coefficients between S-index and line depth, width, and center of mass.
First, we summarize how these correlation coefficients were calculated in W18.
Then we compare these correlation coefficients to spectral line and atomic properties extracted from the Vienna Atomic Line Database (VALD) \citep{kupka99,piskunov95}.

\begin{figure}
    \centering
    \includegraphics{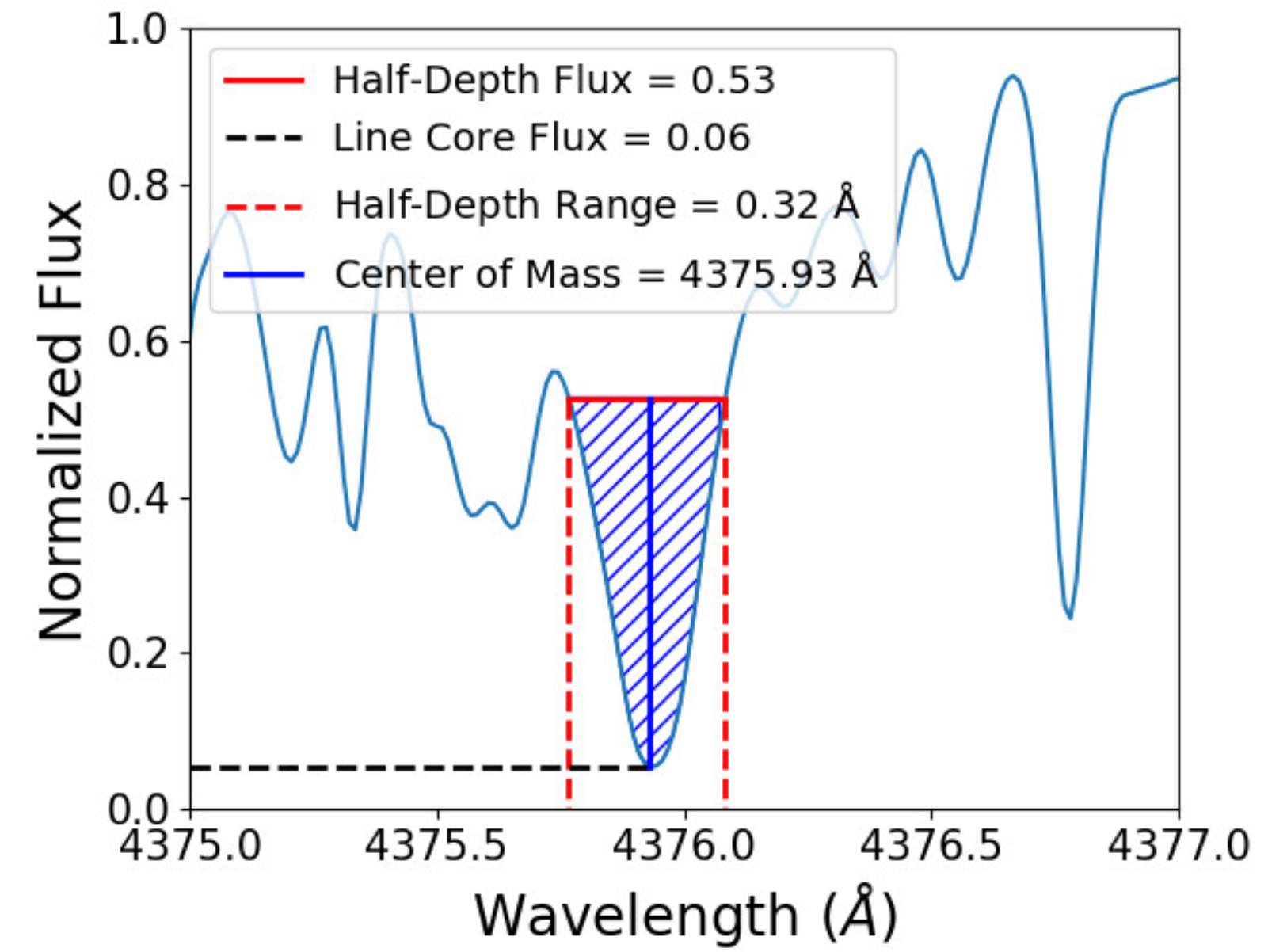}
    \caption{Graphic showing how line properties were measured in W18. The continuum has a normalized flux of 1, and this line has a depth of 0.94. Half-depth flux is a constant value taken from a normalized master spectrum computed as the median of all individual normalized spectra. There are three time-varying measurements. Line core flux is the minimum of a cubic spline interpolation on continuum normalized order-by-order spectra. Half-depth range is the width of the line at half-depth flux. Center of mass is the weighted barycenter in wavelength of the diagonally hatched area.}
    \label{fig:measurements}
\end{figure}

W18 used a data set of HARPS spectra of $\alpha$~Cen~B from March-June 2010.
In W18, after the order-by-order spectra were continuum normalized by fitting a polynomial to a set of local maxima, an automated pipeline was used to identify wavelength intervals that were either stellar atomic spectral lines or blends 
by first selecting all pixels in the normalized spectrum below a value of 0.8, and then grouping selected pixels within $\Delta \lambda / \lambda = 10^{-5}$ (i.e. 3 km/s) of their nearest selected neighbor into wavelength intervals.
Then the line core flux (1 - depth), half-depth range (line width at half-depth flux), and center of mass (area-weighted center below half-depth) were measured for each wavelength interval, e.g. Figure~\ref{fig:measurements}.
Intervals were rejected if their half-depth flux was greater than the flux of the pixels at the interval edges, effectively only keeping intervals with depth $\ga$ 40\%).
Kendall's Tau correlation coefficients \footnote{The tau-b version of Kendall's Tau was calculated in python using the scipy package.} were calculated between each of these three measurements for each line and the S-index.

Our analysis is restricted to the deepest spectral line within each wavelength interval as follows.
For each wavelength interval, spectral lines and their expected depths are obtained using a VALD ``extract stellar" query using stellar parameters for $\alpha$~Cen~B\footnote{The ``extract stellar" star parameters we used were: microturbulence: 1.1~km~s$^{-1}$, $T_{\rm eff}$: 5200~K, $\log{g}$: 4.5~(g in cgs units), and chemical composition: ``Fe: -4.34". The last parameter is VALD's default stellar iron abundance plus 0.2 dex \citep{portodemello08}.}, with a threshold line depth $>$ 0.1.
Out of all of the VALD lines within $2 \times 10^{-5} \times \lambda$ (i.e. 6 km/s), where $\lambda$ is the wavelength of minimum flux within each wavelength interval, the line of maximum VALD depth is treated as the only line within that interval for the extraction of atomic properties from VALD.
We also remove the first 30 out of 72 HARPS orders (approx. 3800 to 4650 \AA) from this analysis due to significant line blending in those orders, and remove lines with excitation energy greater than 8 eV to exclude the Balmer series.
As a result we consider 636 spectral lines in this section, with wavelengths ranging $\sim$4650 to 6900 \AA.

\begin{figure}
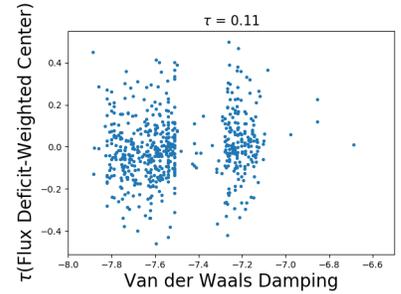

\centering

\gridline{
\fig{"Lande_Factor_vs_flux".png}{0.3\textwidth}{(m) $\tau$(Line Core Flux) vs. Land{\'e} Factor}
\fig{"Lande_Factor_vs_hdr".png}{0.3\textwidth}{(n) $\tau$(Half-Depth Range) vs. Land{\'e} Factor}
\fig{"Lande_Factor_vs_center".png}{0.3\textwidth}{(o) $\tau$(Flux Deficit-weighted Center) vs. Land{\'e} Factor}
}

\gridline{
\fig{"Radiation_Damping_vs_flux".png}{0.3\textwidth}{(p) $\tau$(Line Core Flux) vs. Radiation Damping Coefficient}
\fig{"Radiation_Damping_vs_hdr".png}{0.3\textwidth}{(q) $\tau$(Half-Depth Range) vs. Radiation Damping Coefficient}
\fig{"Radiation_Damping_vs_center".png}{0.3\textwidth}{(r) $\tau$(Flux Deficit-weighted Center) vs. Radiation Damping Coefficient}
}

\gridline{
\fig{"Stark_Damping_vs_flux".png}{0.3\textwidth}{(s) $\tau$(Line Core Flux) vs. Stark Damping Coefficient}
\fig{"Stark_Damping_vs_hdr".png}{0.3\textwidth}{(t) $\tau$(Half-Depth Range) vs. Stark Damping Coefficient}
\fig{"Stark_Damping_vs_center".png}{0.3\textwidth}{(u) $\tau$(Flux Deficit-weighted Center) vs. Stark Damping Coefficient}
}

\gridline{
\fig{"Van_der_Waals_Damping_vs_flux".png}{0.3\textwidth}{(v) $\tau$(Line Core Flux) vs. Van der Waals Damping Coefficient}
\fig{"Van_der_Waals_Damping_vs_hdr".png}{0.3\textwidth}{(w) $\tau$(Half-Depth Range) vs. Van der Waals Damping Coefficient}
\fig{"Van_der_Waals_Damping_vs_center".png}{0.3\textwidth}{(x) $\tau$(Flux Deficit-weighted Center) vs. Van der Waals Damping Coefficient}
}

\caption{Plots visualizing the relationships between each line measurement in W18 and each line property we extract from VALD.
These plots are provided to show why we choose the trend in (g) for deeper analysis. Above each plot is listed a new tau-b value for the correlation between the y-axis tau and the x-axis line property.
\label{fig:correlations}}

\end{figure}

The atomic and spectral properties of each line are retrieved from the VALD query.
These properties are:
\begin{enumerate}
    \item Wavelength in air (\AA).
    \item Line depth: Ratio between line absorption and continuum absorption at the line's wavelength.
    \item Excitation energy: Energy required to reach the lower electronic state of the transition from the ground state (eV).
    \item Oscillator strength: log($gf$) where f is the absorption/emission probability and g is the statistical weight of the lower energy level.
    \item Land{\'e} factor: Factor determining the magnitude of energy level splitting in a weak magnetic field (no units).
    \item Radiation damping coefficient: The rate of decay of the transition, expressed as the logarithm of the radiation damping constant in (4$\pi$s)$^{-1}$.
    \item Stark damping coefficient: Also known as the pressure broadening constant, expressed as the logarithm of the Stark damping constant in (4$\pi$s $N_\mathrm{e}$)$^{-1}$ at 10,000 K
    (where $N_\mathrm{e}$ is the number density of electrons)
    \item Van der Waals damping coefficient: Describes the strength of interaction between neutral atoms, expressed as the logarithm of the Van der Waals damping constant in (4$\pi$s $N_\mathrm{H}$)$^{-1}$ at 10,000 K
    (where $N_\mathrm{H}$ is the number density of neutral Hydrogen)
\end{enumerate}
In Figure~\ref{fig:correlations}, we plot the Kendall's Tau correlation coefficients as in W18 on the vertical axes, and each of these eight properties on the horizontal axes. These taus on the vertical axis represent how activity sensitive each line's depth (line core flux), width (half-depth range), and RV (flux deficit-weighted center) are. Therefore, these plots show how each line's activity sensitivity depends on each on the eight properties above.

Now we look for relationships between these correlation coefficients and the spectral line and atomic properties of each wavelength interval. We quantify the strength of each relationship using Kendall's Tau
\citep[tau-b version, ][]{stuart53}
as the underlying data do not appear to be normally distributed.
Hence we are calculating a new tau, printed above each plot in Figure~\ref{fig:correlations}, between each vertical-axis tau and each horizontal-axis line property.
Figure~\hyperref[fig:correlations]{\ref{fig:correlations}g} has the largest magnitude
of these new
correlation coefficients, $|\tau|$ = 0.41, and by eye appears moderately significant. In the next section, we explore an astrophysical motivation for this trend using a toy model for line depth changes derived from principles of stellar atmospheres.

\begin{figure}
\centering

\gridline{
\fig{"Wavelength_vs_flux".png}{0.3\textwidth}{(a) $\tau$(Line Core Flux) vs. Wavelength}
\fig{"Wavelength_vs_hdr".png}{0.3\textwidth}{(b) $\tau$(Half-Depth Range) vs. Wavelength}
\fig{"Wavelength_vs_center".png}{0.3\textwidth}{(c) $\tau$(Flux Deficit-weighted Center) vs. Wavelength}
}

\gridline{
\fig{"Line_Depth_vs_flux".png}{0.3\textwidth}{(d) $\tau$(Line Core Flux) vs. Line Depth}
\fig{"Line_Depth_vs_hdr".png}{0.3\textwidth}{(e) $\tau$(Half-Depth Range) vs. Line Depth}
\fig{"Line_Depth_vs_center".png}{0.3\textwidth}{(f) $\tau$(Flux Deficit-weighted Center) vs. Line Depth}
}

\gridline{
\fig{"Excitation_Energy_vs_flux".png}{0.3\textwidth}{(g) $\tau$(Line Core Flux) vs. Excitation Energy}
\fig{"Excitation_Energy_vs_hdr".png}{0.3\textwidth}{(h) $\tau$(Half-Depth Range) vs. Excitation Energy}
\fig{"Excitation_Energy_vs_center".png}{0.3\textwidth}{(i) $\tau$(Flux Deficit-weighted Center) vs. Excitation Energy}
}

\gridline{
\fig{"Oscillator_Strength_vs_flux".png}{0.3\textwidth}{(j) $\tau$(Line Core Flux) vs. Oscillator Strength}
\fig{"Oscillator_Strength_vs_hdr".png}{0.3\textwidth}{(k) $\tau$(Half-Depth Range) vs. Oscillator Strength}
\fig{"Oscillator_Strength_vs_center".png}{0.3\textwidth}{(l) $\tau$(Flux Deficit-weighted Center) vs. Oscillator Strength}
}

\end{figure}

\section{Modeling Line Depth Variations} \label{sec:model}

The strongest correlation in the previous section is between electronic excitation energy and the
Kendall's Tau correlation coefficients
between line core flux and S-index, as seen in Figure~\hyperref[fig:correlations]{\ref{fig:correlations}g}. Here we explore that relationship in more detail.
First we describe the analytic stellar atmosphere model that we adopt to explain Figure~\hyperref[fig:correlations]{\ref{fig:correlations}g}.
Then we apply this model, and discuss what we can learn about small variations in the disk-averaged temperature of $\alpha$~Cen~B based on this model and these observations.

\subsection{Stellar Atmosphere Model}

Here we model changes in line depth with excitation energy as arising from changes in disk-averaged temperature, considering changes in absorption and in continuum thermal emission for each line.
For the absorption, we follow the analytic model developed in \cite{gray08}, equation 13.21. We calculate the ratio, $R$, of line absorption, $l_\nu$, to continuous absorption, $\kappa_\nu$:
\begin{equation}
R = \frac{l_\nu}{\kappa_\nu}. \label{eq:R}
\end{equation}
The line absorption $l_\nu$ is proportional to the number of absorbers, i.e.
\begin{equation}
l_\nu = \mathrm{constant\ } e^{- \epsilon / k T} \label{eq:l}
\end{equation}
where $\epsilon$ is the excitation energy of the lower state of the line's transition, $k$ is boltzmann's constant, and $T$ is temperature.
For the visible region of FGK stars, we can model $\kappa_\nu$ as the negative hydrogen ion's bound-free absorption, i.e. 
\begin{equation}
\kappa_\nu = \mathrm{constant\ } T^{-5/2}\ P_\mathrm{e}\ e^{0.75 \mathrm{eV} /kT} \label{eq:k}
\end{equation}
$T$ is temperature, $P_\mathrm{e}$ is the electronic pressure, and 0.75 eV is the binding energy of the negative hydrogen ion.
We approximate the electron pressure as
\begin{equation}
P_\mathrm{e} \approx \mathrm{constant\ } e^{\Omega T} \label{eq:P}
\end{equation}
where $\Omega = 0.0015$ for FGK dwarfs at an optical depth $\tau = 1$ \citep{gray08}. Combining equations \ref{eq:R} -- \ref{eq:P} and taking a natural log gives
\begin{equation}
\mathrm{ln} R = \mathrm{constant} + \frac{5}{2} \mathrm{ln}T - \frac{\epsilon + 0.75 \mathrm{eV}}{kT} - \Omega T.
\end{equation}
This gives us the equation for the change in line depth with temperature due to absorption:
\begin{equation}
\frac{1}{R} \dv{R}{T} = \frac{2.5}{T} + \frac{\epsilon + 0.75 \mathrm{eV}}{kT^2} - \Omega.
\end{equation}
Next we model the continuum emission using the Planck function
\begin{equation}
B_\nu = \frac{2 h \nu^3}{c^2} \frac{1}{e^{h \nu / k T} - 1}.
\end{equation}
The fractional rate of change in emission with temperature is
\begin{equation}
\frac{1}{B_\nu} \dv{B_\nu}{T} = \frac{h \nu}{k T^2} \frac{1}{1 - e^{-h \nu / k T}}.
\end{equation}
Putting together the equations for continuum thermal emission and line absorption, we obtain the model of how a line core's normalized flux, $F$, changes with temperature:
\begin{equation}
\dv{F}{T} = \frac{1}{B_\nu} \dv{B_\nu}{T} - \dv{R}{T} = \frac{h \nu}{k T^2} \frac{1}{1 - e^{-h \nu / k T}} - R \left[ \frac{2.5}{T} + \frac{\epsilon + 0.75 \mathrm{eV}}{kT^2} - \Omega \right].
\end{equation}

In the middle part of this equation, the term $dR/dT$ does not have a $1/R$ factor (as the $B_\nu$ term does) since it is already normalized by the absorption of the continuum in Eq. 1. For the right side of the equation, 
\cite{gray08} suggest using a temperature about 15\% below the effective temperature as characteristic of the line forming region, so starting with $T_\mathrm{eff} = 5230$ K for $\alpha$~Cen~B \citep{brandenburg17} gives us $T = 4445.5$ K.
Note that we also tried using $T = 5230$ in the model, and our resulting $\Delta T$ in Section~\ref{fittingdata} below changed from 3.3 K to 3.6 K, a $<$ 10\% difference in our result.
In applying our model, we set the VALD values for line depth equal to $R$ in Eq.~9, as our formulation of $R$ is a good approximation for line depth for ``weak lines of a neutral species with the element mostly neutral" as shown in \cite{gray08}, which are the types of lines we consider below.
Next we use this model, along with the results of W18, to learn about disk-averaged temperature variations in $\alpha$~Cen~B.

\subsection{Fitting the Data}\label{fittingdata}
Here we apply the above line depth model to the line depth measurements from W18. To simplify the application of the model, we first apply the following two filters to our list of 636 spectral used in Section~\ref{sec:correlations}. First, we remove all blends where the range in excitation energies of spectral lines in the blend is greater than 1 eV. This eliminates ``line mislabeling" cases where a shallow, low-excitation line dominates the line depth variations over a deeper line in a blend. Then we remove all spectral lines other than Fe I lines, as Fe I lines account for a large fraction of lines in the visible spectrum and are primarily formed in the stellar photosphere. After these two filters, 236 iron lines remain. We remove eight additional lines that are clearly contaminated by tellurics, as described in the next section. As a result, 228 iron lines are used the following analysis.

\begin{figure}
\centering
\includegraphics[width=0.9\linewidth]{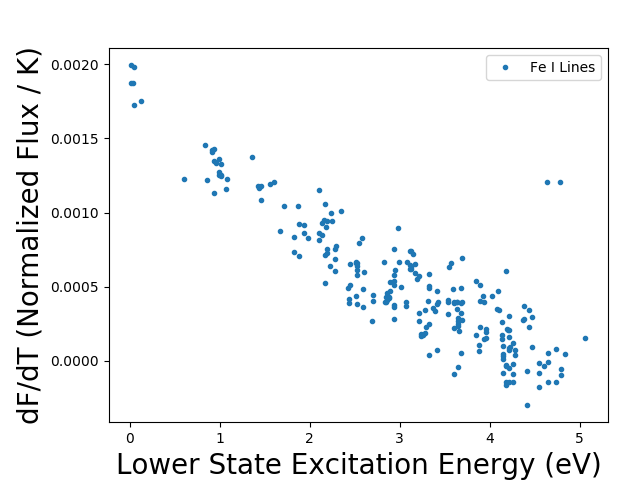}
\caption{The change in line core flux with temperature (d$F$/d$T$) for the model is plotted versus the lower state excitation energy. The two iron lines near the coordinates (4.8,0.0012) are distinct because of their small VALD depths ($\sim$0.15). \label{fig:model}}
\end{figure}

\begin{figure}
\centering
\includegraphics[width=0.9\linewidth]{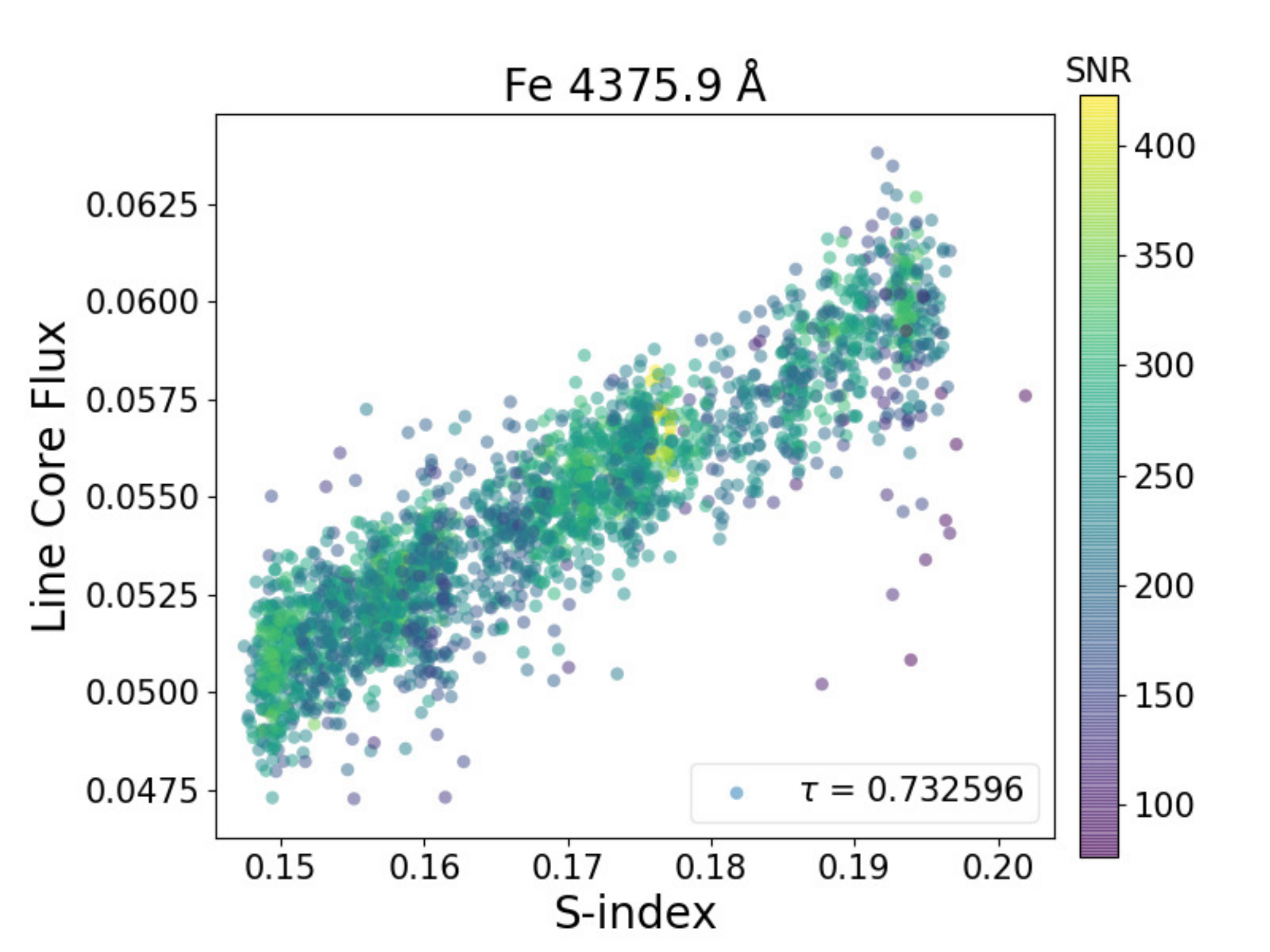}
\caption{An example correlation plot from W18.
This plot demonstrates that a linear fit is a good approximation for the relationship between S-index and line core flux.} \label{fig:correlation-example}
\end{figure}

\begin{figure}
\centering
\includegraphics[width=0.9\linewidth]{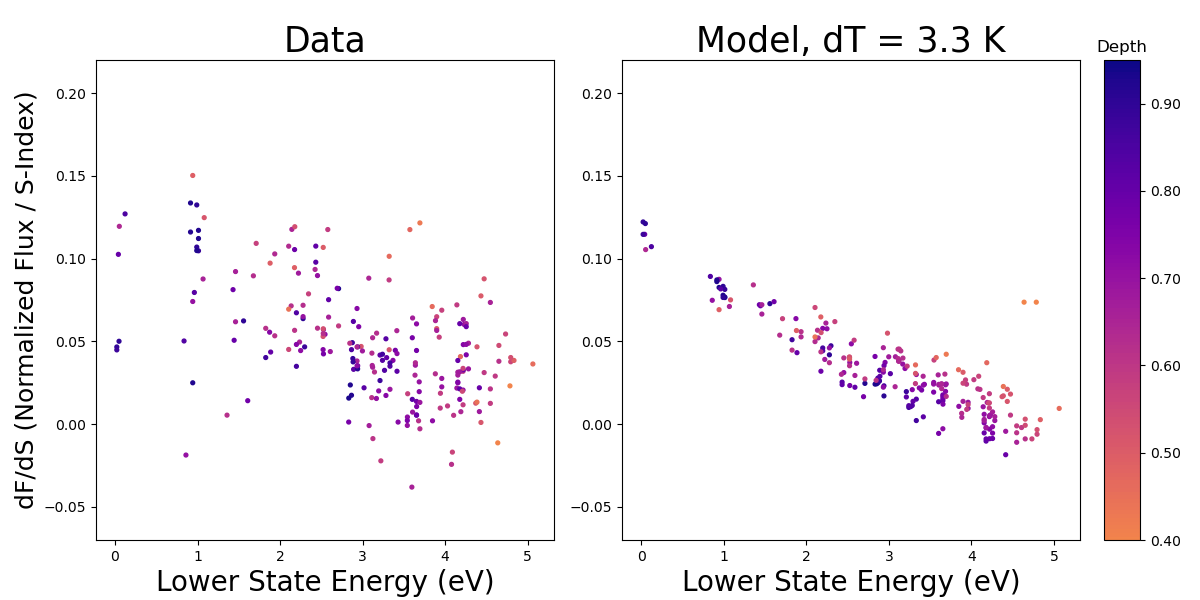}
\caption{Comparison of our measured values of d$F$/d$S$ on the vertical axis (left panel) to our best-fit model for d$F$/d$S$ (right panel), both as a function of lower state excitation energy. The color scale shows line depth and is truncated at 0.5 to improve visualization for the majority of data points. The two model points near the coordinates (4.8,0.07) are distinct because of their very small VALD line depths ($\sim$0.15).  Note, that the vertical axis on the right panel is not the same as shown in Figure \ref{fig:model}, but is scaled by a constant d$T$/d$S$. \label{fig:dataModelComparison}}
\end{figure}

Figure~\ref{fig:model} shows the result of our model calculations of d$F$/d$T$ for all of the iron lines in our sample. We see increased temperature sensitivity at lower excitation potential as in previous models and solar observations \citep[e.g.][]{rutten1983, elste1986, takeda2017}. This model tells us how line core flux changes with temperature, while W18 explored how line core flux changes with S-index, $S$ \citep[e.g. as in][]{brandt1990}.
To fit the model to the line core flux measurements for $\alpha$~Cen~B, we consider the equation
\begin{equation}
\dv{F}{S} = \dv{T}{S} \dv{F}{T}. \label{eq:dFdS}
\end{equation}
We calculate d$F$/d$S$ using the line core flux and S-index measurements in W18.
Looking at the correlation plots in W18, an example of which is shown in Figure~\ref{fig:correlation-example}, it seems a linear fit is a good approximation for the relationship between S-index and line core flux.
Hence we use a weighted least squares regression to find the slope of the line of best-fit for each Fe line in our sample.
The weights are the signal-to-noise ratios (SNRs) per pixel, calculated by taking the average over the middle 200 pixels of order 35 in each spectrum.
We plot these slopes on the vertical axis of Figure~\ref{fig:dataModelComparison}-left.

The only term we have not calculated in Eq.~\ref{eq:dFdS} is d$T$/d$S$, the change in temperature with S-index.
We assume this is a constant within the range of S-index values for this data set, as this is the simplest approximation we can make and any consideration of a more complicated relationship is beyond the scope of this work. One consequence of this assumption is that Figures \ref{fig:model} and \ref{fig:dataModelComparison}-right appear identical (except for different y-axis scales). d$T$/d$S$ is calculated by dividing the slopes of best-fit lines to Figures \ref{fig:model} and \ref{fig:dataModelComparison}-left.

Lastly, we calculate the temperature change between the more ``active" hemisphere (rotational phase with highest S-index) and ``quiet" hemisphere (lowest S-index) of the star using
\begin{equation}
\Delta T = \dv{T}{S} \Delta S
\end{equation}
where $\Delta S$ is the full range of S-index values in our data.
To aid in the interpretation of this $\Delta T$, we considered modifying the model for d$F$/d$T$ to include a constant multiplying the whole equation, $A$, which we interpret as the fractional active area on the disk.  However, $A$ was found to be degenerate with the temperature according to
\begin{equation}
A\ \Delta T \approx \mathrm{constant}.
\end{equation}
Thus, our model cannot distinguish between a 300 K hotter active region covering 1\% of the disk and a 3 K hotter active region covering 100\% of the disk. Consequently, we interpret the temperature $T$ as the disk-averaged temperature.

\begin{figure}
\centering
\includegraphics[width=0.9\linewidth]{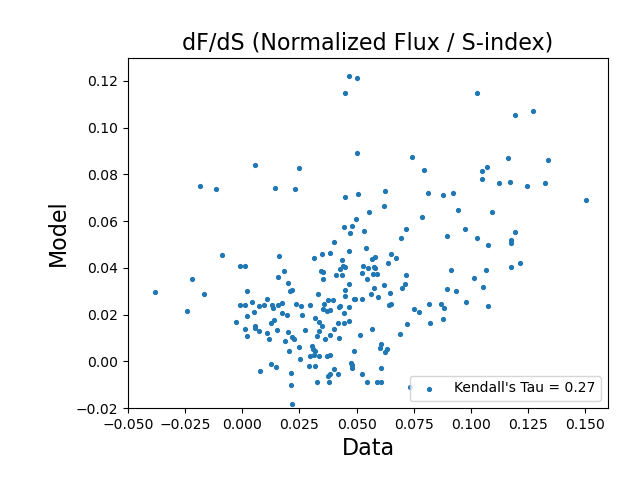}
\caption{Correlation plot of our best-fit model vs. measured values of d$F$/d$S$. \label{fig:dataModelCorrelation}}
\end{figure}

Figure~\ref{fig:dataModelComparison} shows a side-by-side comparison of our measured values of d$F$/d$S$ with our model for d$F$/d$S$ calculated using Eqs. 9 and 10. Note that because we calculated d$T$/d$S$ by dividing the slopes of best-fit lines to Figures \ref{fig:model} and \ref{fig:dataModelComparison}-left, the slope of the right-hand panel of Figure~\ref{fig:dataModelComparison} has been artificially matched to the slope of the left-hand panel. However, this process does not align the vertical intercepts of the two plots (value of d$F$/d$S$ when lower state energy = 0). Therefore, the qualitative agreement of the intercepts between the two panels is encouraging.
We also color-code Figure~\ref{fig:dataModelComparison} based on line depth to show the depth-dependence in the model.
Figure~\ref{fig:dataModelCorrelation} shows the correlation plot for the model and data.
In the next section, we discuss possible reasons for the differences between the model and data.

\subsection{Discussion}

We have found a simple stellar atmospheric model of line formation helps explain why some lines have depth variations that are more correlated with S-index than others. As the values of d$F$/d$S$ calculated from the data have a much larger scatter within a small bin of excitation energy than the model values, as seen in Figure~\ref{fig:dataModelComparison}, here we discuss possible sources of error.
Sources of error include blending of spectral lines, which can affect how each line's core depth changes with excitation energy if the line is blended with a line of differing excitation energy, or of a different species.
While we attempt to mitigate the effects of line blending by only choosing blends where the deepest line is Fe I and the range in excitation energy is $<$ 1 eV, we make no attempt model these effects.

Another source of error in this analysis is telluric contamination. This is not an issue in W18 as they find no evidence of telluric contamination for correlation coefficients $\tau > 0.5$. However, the full range of $\tau$ is considered here, so an attempt is made to remove outliers that are most clearly caused by telluric lines.

\begin{figure}
\centering
\includegraphics[width=0.45\linewidth]{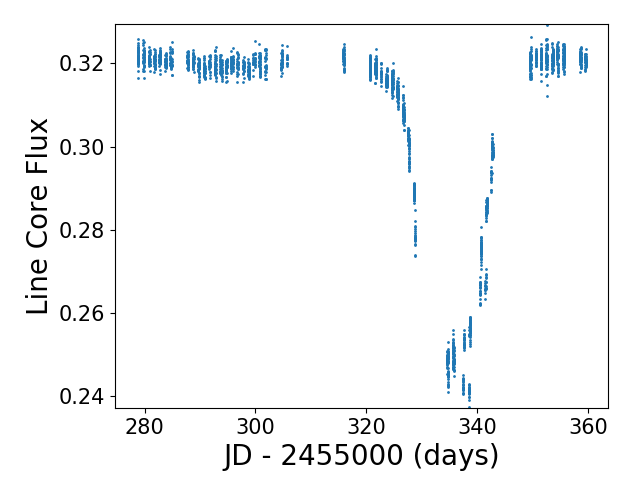}
\includegraphics[width=0.45\linewidth]{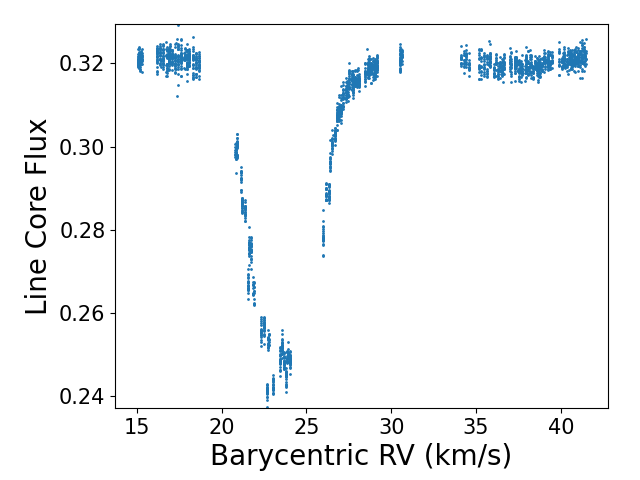}
\caption{Plots showing the effect of a telluric line on the flux measured in the core of a Fe I line. Left: Line core flux plotted against time. As the telluric line passes through the Fe I line's core over time, the Fe 1 line appears deeper between JD-2455000 of 320 and 350.
Right: Line core flux plotted against barycentric RV. The telluric line and stellar line only ``line up" for a small window of RVs, and the full width at half maximum (FWHM) of that window matches the telluric line's $\sim$4 km/s FWHM.
\label{fig:telluric}}
\end{figure}

To flag lines with clear evidence for telluric contamination, first we look at time-series plots of line core flux to identify candidates for telluric contamination, e.g. Figure~\ref{fig:telluric}-Left. To confirm these are tellurics, we then created and watched videos of each candidate spectral line, where the telluric can be seen gradually moving into and/or out of the line due to the annual variation of the barycentric Earth RV. Thirty iron lines that have the largest y-coordinate difference between the left- and right-hand sides of Figure~\ref{fig:dataModelComparison} are considered for this analysis. Eight out of these thirty lines are found to be contaminated by tellurics, and all eight come from the eleven largest differences, so we do not consider additional lines after these thirty.
In the future, an automated approach could be used that considers all lines and measures the velocity range over which the line core flux variations occur, as shown in Figure~\ref{fig:telluric}-Right. Comparing this velocity range to the typical telluric line's velocity width could help discriminate between telluric contamination and other signals.

If we had not removed these eight telluric-contaminated lines, several of them would correspond to data points outside of the vertical axes of Figure~\ref{fig:dataModelComparison}-left. Now, there are no data points outside of these axes. Therefore we suggest that some of the remaining scatter is due to telluric lines that are too weak to be identified by the visual analysis described above.

The following are also suspected to be causes of error in our d$F$/d$S$ measurements, though we were unable to find convincing evidence of these effects.
Iron lines at bluer wavelengths may have weaker responses to temperature than predicted by the model due to an incorrect line formation temperature (we estimated this as 85\% of $T_\mathrm{eff}$ for all lines). This issue may have been mitigated by discarding the first 30 out of 72 HARPS orders.
Also, poor continuum normalizations in line-blanketed orders or higher photon noise at the bluer orders may contribute to the scatter in our measurements of d$F$/d$S$. In the next section we summarize the results of this paper and discuss future avenues of research.

\section{Conclusions and Future Work} \label{sec:conclusion}

We searched for correlations between the atomic properties of spectral lines and the correlation coefficients between S-index and line depth, width and barycenter. W18 derived these S-index correlation coefficients using HARPS spectra of $\alpha$~Cen~B during a time of high, sinusoidally varying activity. We identify a particularly interesting correlation between the line core flux correlation coefficients and the lower state excitation energy, a correlation not previously identified. This correlation demonstrates that (Fe I) lines with smaller excitation energy have more activity-sensitive depths.

We then developed a toy model of the expected line core flux changes with temperature that qualitatively reproduces the observations. This model can be used to estimate the temperature difference in magnetically active regions. The line core flux changes in this particular data set could be explained by the disk-averaged temperature being 3.3~K hotter when the star shows maximum activity, compared to minimum activity. In \cite{dumusque14}, the author shows that during the time of the data analyzed here, a
facula
of 2.12$^{+0.88}_{-0.56}$ \% of the visible hemisphere is covering
part of
the stellar surface. Linking the 3.3 K average increase in temperature and the 2.12 \% size of the active region, we can estimate an active region temperature contrast of
150~K
hotter than the quiet stellar surface
using the Stephan-Boltzmann law\footnote{$T_\mathrm{faculae}$ is calculated from the Stephen-Boltzmann law by setting the fluxes equal: $(T_\sun+3.3~\mathrm{K})^4 = (1-0.0212)*T_\sun^4 + 0.0212*T_\mathrm{active}^4$ where $T_\sun$ = 5778~K}
This implies the presence of a large or several small faculae on the surface of $\alpha$~Cen~B. The temperature contrast is within the large expected range \citep[e.g. 34~K to 251~K from disk center to edge in][]{meunier10}.

As we improve our models of how stellar activity affects RV measurements, we would like to understand the dominant ways in which stellar activity affects the spectrum on a physical basis. The present study can be seen as a step toward physically motivated models of how stellar activity affects every pixel in a spectrum. W18 found that line core flux measurements had the strongest correlations with S-index, and now we can 
partially
understand those correlations using a simple analytic model of stellar atmospheric absorption and emission.
As line depth does not affect CCF-based RV measurements, except in cases of blended lines or template errors, this line core flux correlation is distinct from traditional measurements of line shape changes such as bisector span and chromatic index \citep[e.g.][]{queloz2001,zechmeister2018}. Our analysis provides a ``new'' way of looking at how the spectra of stars due to activity are dynamic, and offers an ``old'' astrophysical explanation thereof. In the future, we aim
to continue to assign physical explanations to the ways in which stellar activity affects RV spectra, 
including the various measurements of line shape changes.

In traditional Doppler RV analysis, the spectrum of a star is taken to be static in time. However, we have shown that stellar spectra of interest to planet hunters are measurably dynamic in response to changes in stellar activity level. \cite{rajpaul20} propose a new
data-driven
method of measuring RVs that applies this principle by not requiring a static stellar template. In future work, we plan to take a look at how changing line depths can induce RV measurement error due to the mask no longer correctly weighting each line. The path forward requires treating the stellar spectrum as dynamic rather than static.

\acknowledgments

This research is based on data products from observations made with ESO Telescopes at the La Silla Paranal Observatory under programme IDs 60.A-9036(A), 072.C-0488(E), 072.C-0513(B), 072.C-0513(D), 074.C-0012(A), 074.C-0012(B), 076.C-0878(A), 077.C-0530(A), 078.C-0833(A), 079.C-0681(A), 084.C-0229(A), 085.C-0318(A), and 192.C-0852(A).
This research has made use of the services of the ESO Science Archive Facility, and is based on data obtained from the ESO Science Archive Facility under request numbers 243402, 243835, 300034, 300050. This work has made use of the VALD database, operated at Uppsala University, the Institute of Astronomy RAS in Moscow, and the University of Vienna. A. W. acknowledges support from the Delaware Space Grant College and Fellowship Program, NASA Grant NNX15AI19H.
This work was partially supported by funding from the Center for Exoplanets and Habitable Worlds. The Center for Exoplanets and Habitable Worlds is supported by the Pennsylvania State University and the Eberly College of Science. This research was partially supported by Heising-Simons Foundation Grant \#2019-1177.
C X. D. acknowledges support from the Branco-Weiss--Society in Science fellowship.
This project has received funding from the European Research Council (ERC) under the European Union Horizon 2020 research and innovation program (grant agreement No. 851555).
P.P. acknowledges support from NASA (award 16-APROBES16-0020 and support from the Exoplanet Exploration Program) and the National Science Foundation (Astronomy and Astrophysics grant 1716202), the Mount Cuba Astronomical Foundation and George Mason University start-up funds.
H.M.C acknowledges financial support from the National Centre for Competence in Research (NCCR) PlanetS, supported by the Swiss National Science Foundation (SNSF),
as well as a UK Research and Innovation Future Leaders Fellowship.

\clearpage


\begin{thebibliography}{}


\bibitem[Baliunas et al.(1995)]{baliunas95} Baliunas, S. L., Donahue, R. A., Soon, W. H., et al.\ 1995, \apj, 438, 269

\bibitem[Blackman et al.(2020)]{blackman20} Blackman, R.~T., Fischer, D.~A., Jurgenson, C.~A., et al.\ 2020, \aj, 159, 238

\bibitem[Brandenburg et al.(2017)]{brandenburg17} Brandenburg, A., Mathur, S., \& Metcalfe, T.~S.\ 2017, \apj, 845, 79

\bibitem[Brandt \& Solanki(1990)]{brandt1990} Brandt, P.~N. \& Solanki, S.~K.\ 1990, \aap, 231, 221

\bibitem[Brandt \& Getling(2008)]{brandt08} Brandt, P.~N., \& Getling, A.~V.\ 2008, \solphys, 249, 307


\bibitem[Collier Cameron et al.(2021)]{collier21} Collier Cameron, A., Ford, E.~B., Shahaf, S., et al.\ 2021, \mnras, 505, 1699. doi:10.1093/mnras/stab1323

\bibitem[Cretignier et al.(2020)]{cretignier20} Cretignier, M., Dumusque, X., Allart, R., et al.\ 2020, \aap, 633, A76

\bibitem[Cretignier et al.(2021)]{cretignier2021} Cretignier, M., Dumusque, X., Hara, N.~C., et al.\ 2021, arXiv:2106.07301


\bibitem[de Beurs et al.(2020)]{deBeurs20} de Beurs, Z.~L., Vanderburg, A., Shallue, C.~J., et al.\ 2020, arXiv:2011.00003

\bibitem[De Rosa \& Toomre(2004)]{derosa04} De Rosa, M.~L., \& Toomre, J.\ 2004, \apj, 616, 1242

\bibitem[Dumusque(2014)]{dumusque14} Dumusque, X.\ 2014, \apj, 796, 133

\bibitem[Dumusque(2018)]{dumusque18} Dumusque, X.\ 2018, \aap, 620, A47

\bibitem[Dumusque et al.(2012)]{dumusque12} Dumusque, X., Pepe, F., Lovis, C., et al.\ 2012, \nat, 491, 207

\bibitem[Dumusque et al.(2011)]{dumusque11} Dumusque, X., Santos, N.~C., Udry, S., Lovis, C., \& Bonfils, X.\ 2011, \aap, 527, A82


\bibitem[Elste(1986)]{elste1986} Elste, G.\ 1986, \solphys, 107, 47. doi:10.1007/BF00155340


\bibitem[Gray(2008)]{gray08} Gray, D.~F.\ 2008, The Observation and Analysis of Stellar Photospheres


\bibitem[Haywood et al.(2020)]{haywood20} Haywood, R.~D., Milbourne, T.~W., Saar, S.~H., et al.\ 2020, arXiv:2005.13386


\bibitem[Jurgenson et al.(2016)]{jurgenson16} Jurgenson, C., Fischer, D., McCracken, T., et al.\ 2016, \procspie, 9908, 99086T


\bibitem[Kupka et al.(1999)]{kupka99} Kupka, F., Piskunov, N., Ryabchikova, T.~A., et al.\ 1999, \aaps, 138, 119. doi:10.1051/aas:1999267


\bibitem[Meunier et al.(2010)]{meunier10} Meunier, N., Desort, M., \& Lagrange, A.-M.\ 2010, \aap, 512, A39

\bibitem[Miklos et al.(2020)]{miklos20} Miklos, M., Milbourne, T.~W., Haywood, R.~D., et al.\ 2020, \apj, 888, 117. doi:10.3847/1538-4357/ab59d5


\bibitem[Pepe et al.(2013)]{pepe13} Pepe, F., Cristiani, S., Rebolo, R., et al.\ 2013, The Messenger, 153, 6

\bibitem[Piskunov et al.(1995)]{piskunov95} Piskunov, N.~E., Kupka, F., Ryabchikova, T.~A., et al.\ 1995, Laboratory and Astronomical High Resolution Spectra, 81, 610

\bibitem[Plavchan et al.(2020)]{plavchan20} Plavchan, P., Vasisht, G., Beichman, C., et al.\ 2020, arXiv e-prints, arXiv:2006.13428

\bibitem[Porto de Mello et al.(2008)]{portodemello08} Porto de Mello, G.~F., Lyra, W., \& Keller, G.~R.\ 2008, \aap, 488, 653


\bibitem[Queloz et al.(2001)]{queloz2001} Queloz, D., Henry, G.~W., Sivan, J.~P., et al.\ 2001, \aap, 379, 279. doi:10.1051/0004-6361:20011308


\bibitem[Rajpaul et al.(2020)]{rajpaul20} Rajpaul, V.~M., Aigrain, S., \& Buchhave, L.~A.\ 2020, \mnras, 492, 3960

\bibitem[Rutten \& Zwaan(1983)]{rutten1983} Rutten, R.~J. \& Zwaan, C.\ 1983, \aap, 117, 21


\bibitem[Saar \& Donahue(1997)]{saar97} Saar, S. H., \& Donahue, R. A.\ 1997, ApJ, 485, 319

\bibitem[Stuart, A. (1953)]{stuart53} Stuart, A.\ 1953, Biometrika, 40(1/2), 105. https://doi.org/10.2307/2333101

\bibitem[Su{\'a}rez Mascare{\~n}o et al.(2020)]{suarez20} Su{\'a}rez Mascare{\~n}o, A., Faria, J.~P., Figueira, P., et al.\ 2020, arXiv e-prints, arXiv:2005.12114


\bibitem[Takeda \& UeNo(2017)]{takeda2017} Takeda, Y. \& UeNo, S.\ 2017, \solphys, 292, 123. doi:10.1007/s11207-017-1144-x

\bibitem[Thompson et al.(2017)]{thompson17} Thompson, A.~P.~G., Watson, C.~A., de Mooij, E.~J.~W., et al.\ 2017, \mnras, 468, L16

\bibitem[Thompson et al.(2020)]{thompson20} Thompson, A.~P.~G., Watson, C.~A., Haywood, R.~D., et al.\ 2020, \mnras, 494, 4279


\bibitem[Wilson(1978)]{wilson78} Wilson, O.~C.\ 1978, \apj, 226, 379

\bibitem[Wise et al.(2018)]{wise18} Wise, A.~W., Dodson-Robinson, S.~E., Bevenour, K., et al.\ 2018, \aj, 156, 180

\bibitem[Wright(2017)]{wright17} Wright, J.~T.\ 2017, Handbook of Exoplanets, Edited by Hans J.~Deeg and Juan Antonio Belmonte.~Springer Living Reference Work, ISBN: 978-3-319-30648-3, 2017, id.4, 4


\bibitem[Zechmeister et al.(2018)]{zechmeister2018} Zechmeister, M., Reiners, A., Amado, P.~J., et al.\ 2018, \aap, 609, A12. doi:10.1051/0004-6361/201731483

\end{thebibliography}
\end{document}